# Bistable perception, precision and neuromodulation

Filip Novicky[1,2,] *, Thomas Parr[3], Karl Friston[3], M. Berk Mirza[4], Noor Sajid[3]

[1] Department of Neurophysics, Radboud University, Nijmegen, Netherlands.
[2] Maastricht University, Maastricht, Netherlands.
[3] Wellcome Centre for Human Neuroimaging, UCL, London, United Kingdom.
[4] Department of Psychology, University of Cambridge, Cambridge, United Kingdom.

*Correspondence to: filip.novicky@donders.ru.nl

## Abstract

Bistable perception follows from observing a static, ambiguous, (visual) stimulus with two possible interpretations. Here, we present an active (Bayesian) inference account of bistable perception and posit that perceptual transitions between different interpretations (i.e., inferences) of the same stimulus ensue from specific eye movements that shift the focus to a different visual feature. Formally, these inferences are a consequence of precision control that determines how confident beliefs are and change the frequency with which one can perceive – and alternate between – two distinct percepts. We hypothesised that there are multiple, but distinct, ways in which precision modulation can interact to give rise to a similar frequency of bistable perception. We validated this using numerical simulations of the Necker's cube paradigm and demonstrate the multiple routes that underwrite the frequency of perceptual alternation. Our results provide an (enactive) computational account of the intricate precision balance underwriting bistable perception. Importantly, these precision parameters can be considered the computational homologues of particular neurotransmitters – i.e., acetylcholine, noradrenaline, dopamine – that have been previously implicated in controlling bistable perception, providing a computational link between the neurochemistry and perception.

# Introduction

Bistable perception ensues from observing a static, ambiguous stimulus with two possible interpretations e.g., the Necker cube or Rubin's vase. Here, alternation of the visual percept arises when the stimulus offers two distinct explanations that cannot be perceived simultaneously (Brascamp et al., 2018). For example, whilst observing Rubin's vase, individuals switch between perceiving a black vase or two facial profiles. Experimentally, it has been shown that neurotransmitters are crucial for modulating this phenomenon (van Loon et al., 2013) – specifically, implicating catecholaminergic (Pfeffer et al., 2018), dopaminergic (Schmack et al., 2013), cholinergic (Sheynin et al., 2020), and noradrenergic (Einhauser et al., 2008)[1] neurotransmission in modulating the frequency of perceptual switching. In this study, we provide a computational account of how these particular neurotransmitters can influence bistable perception. For this, we rely on how their computational homologues – i.e., precision modulation under active (Bayesian) inference (Parr & Friston, 2017) – can induce perceptual alternation.

Active inference is a Bayesian formulation of brain function that casts perception and action as 'self-evidencing' (Hohwy, 2016); or minimising free energy across time (Da Costa et al., 2020; Friston, 2019; Friston et al., 2017; Kaplan & Friston, 2018). It characterises perception as an inferential process (Clark, 2013), across the space of all possible hypotheses that could have given rise to a particular stimulus (Friston, 2005). These inferences are a consequence of how confident (or precise) beliefs are over particular model distributions. Broadly, such models comprise sequences of 'hidden' states or causes which generate observable sensory data. For example, if the probability of a sensory input given its cause is extremely precise, then one can confidently attribute that sensory observation to a particular cause. Contrariwise, an imprecise probability distribution implies an ambiguous association between cause and effect and sensory observations can do little to resolve the uncertainty about their causes. This is precisely why precision control can influence the type of inferences made and induce bistable perception by mimicking the role of specific neuromodulators (Moran et al., 2013; Parr et al., 2018; Schwartenbeck et al., 2015; Vincent et al., 2019).

Here, we use particular precision parameters to investigate the computational mechanisms that underwrite bistable perception. We hypothesised that there are multiple, but distinct, ways in which precision control can interact to give rise to bistable perception. These precision manipulations influence the frequency with which one can perceive (and alternate between) two distinct percepts and speak to an intricate precision balance underwriting bistable perception. Explicitly, we evaluate multiple combinations of precision, over three distinct model distributions, that may give rise to bistable perception. These are i) sensory precision, ii) precision over state transitions, and iii) precision over probable action plans, as these are thought to be mediated by acetylcholine (Moran et al., 2013; Parr et al., 2018), noradrenaline (Vincent et al., 2019), and dopamine (Schwartenbeck et al., 2015), respectively.

To demonstrate perceptual switching — as a function of various precisions — we instantiate an active inference model[2] of the Necker cube paradigm (Gregory, 1980). In this example, the agent is presented with an ambiguous, static image, i.e., the Necker cube, and infers its cause; namely, a cube facing either to the right or left. How quickly and often the agent alternates between the two inferred (i.e., perceived) orientations is determined by the confidence with which particular beliefs are updated – modulated by the different precision parameters. We discuss the correspondence between these precision terms, their neuromodulatory homologues and role in facilitating bistable perception in Table 1. Inevitably, these associates are vast oversimplifications. However, they are useful heuristics that appear to be consistent with much of the data on neuromodulatory function.

---

[1] Studied indirectly via pupil dilation – see Larsen, R. S., & Waters, J. (2018). Neuromodulatory Correlates of Pupil Dilation [Mini Review]. *Frontiers in Neural Circuits*, *12*(21). https://doi.org/10.3389/fncir.2018.00021 .
[2] Here, we will use agent and model to mean one and the same thing.

Briefly, sensory precision (i.e., the likelihood function) determines the confidence in beliefs about causes of outcomes and can be associated with (selective) attention (Mirza et al., 2019). Similarly, precision over state transitions models the volatility of hidden states. If this is extremely precise, the agent would have high confidence about the evolution of states over time. Conversely, with a low state transition precision, the agent's beliefs about future states would become progressively more uncertain (i.e., high Shannon entropy). Lastly, the precision over probable action plans (i.e., policy selection) determines the confidence in the selected action trajectory, or policy. We expected that increasing each of these precisions would decrease the frequency of visual perception alternation induced by precise beliefs over the perceived orientation (or the visual context), independently of the other precision terms. Since all precision terms were hypothesised to induce similar consequence on switching rate (see Table 1), we analysed posterior probability of the cube's orientation after the switch occurred to provide a dissociable account of these precision manipulations. In other words, the differential effects of the precision manipulations were assessed in terms of what the synthetic subject 'believed' at the time of each perceptual switch.

Table 1. Overview of precision parameters, and how they may affect bistable perception.

| Precision parameter | Neuromodulators | Bistable perception effects | Computational role |
|---|---|---|---|
| **Sensory** ($\zeta$) | Cholinergic | (Sheynin et al., 2020): increased acetylcholine during bistable perception increased the visibility of individual percepts and decreased the frequency of perceptual transition. (Pfeffer et al., 2018): found no effect of cholinergic release during perception transitions. | Sensory precision positively correlates with the visual accuracy of perceived orientation and negatively with the switch frequency. |
| **State transition** ($\omega$) | Noradrenergic | Catecholamines (i.e., a mixture of dopamine and noradrenaline) negatively correlates with the duration of holding one percept during a multi-stability task (Pfeffer et al., 2018) | State transition precision modulates the evolution of a perceived orientation, and precise state transitions reduce switch frequency. |
| **Policy** ($\gamma$) | Dopaminergic | (Schmack et al., 2013): DRD4-2R (gene) that targets dopaminergic release also influences perceptual switches. However, DRD4-4R and -7R do not show any modulatory effects. | Policy precision is linked to confidence about actions (i.e., eye movements) and can decrease the switch frequency. |

This paper is structured as follows. First, we review formal (i.e., computational) accounts of bistable perception. Next, we briefly introduce active inference with a special focus on precision. This provides a nice segway to introduce our generative model for simulating bistable perception of the Necker's cube; a canonical paradigm in the bistable perception literature (Choi et al., 2020; Kornmeier & Bach, 2005; Wernery et al., 2015). The model is then used to test our hypotheses regarding the multiple, and distinct routes through which bistable perception can arise. Finally, we discuss the results to understand how our simulated manipulations of precision relates to neuromodulation in the brain.

# Computational accounts of bistable perception

Previously, there have been many attempts to account for bistable perception phenomena ranging from dynamical systems models (Fürstenau, 2010, 2014) through to predictive processing frameworks (Hohwy et al., 2008). The latter explanation include a formulation (Weilnhammer et al., 2017) that characterise perceptual switches as a consequence of prediction errors emerging from residual evidence for the suppressed percept. In this account, bistable perception emerges from a progressive increase of the prediction error not explained by the extant percept, engendering the alternate explanation. We extend this account of bistable perception using active inference. Explicitly, we illustrate that variations in precision (over distinct model parameters) can give rise to bistable perception by influencing how confidently sensory observations are inferred.

Our account is also aligned with another model of bistable perception introduced by (Weilnhammer et al., 2021). They observed that bistable perception emerged from a fluctuation in the sensory information available to the brain. This fluctuation can be explained by saccadic suppression—the suppression of sensory pathways during saccades (Crevecoeur & Kording, 2017)—and can lead to increased perceptual alternation. Under predictive processing accounts, this suppression relies upon changes in the precision the brain assigns to sensory data at different times during the action-perception cycle. This highlights that eye movements are necessary for understanding bistable perception and can therefore provide behavioural evidence of sensory precision modulations. Importantly, this aligns with our model by demonstrating that bistable perception is i) modulated via different levels of (sensory) precision and ii) experimentally linked to eye movements; namely, active vision or inference.

Separately, (Parr et al., 2019) used active inference to investigate the computational mechanisms that underwrite bistable perception. They postulated that bistable perception is a consequence of alternations in (covert) attentional deployment towards certain stimulus features when two different percepts may be supported by different stimulus features (e.g., luminance contrast at different places in the visual field). The alternation is a consequence of accumulation of uncertainty about the percept relating to the unattended features. By choosing to deploy attention to resolve this uncertainty, we switch our focus and therefore our percept. The numerical experiments accompanying this hypothesis showed that changes in different precision-parameters influenced the frequency of transitions, given the inferences being made. This process has been linked to eye movements focusing on distinct parts of the illusory object which is in line with a call for active vision formulations of bistable perception (Safavi & Dayan, 2022).

# Bistable perception, precision modulation and active inference

Here, we briefly describe active inference and precision parameters that underwrite the computational mechanisms that may give rise to bistable perception.

### Active inference

Active inference, a corollary of the free energy principle, is a formal way to describe the behaviour of self-organising (random dynamical) systems that exchange with an external environment. It postulates that these systems self-organise by minimising their surprisal about sensory observations[3] ($o$), i.e., maximising their (Bayesian) model evidence (Friston et al., 2010; Sajid, Da Costa, et al., 2021) or 'self-evidencing'(Hohwy, 2016). Formally, this involves the optimisation of a free energy functional i.e., an upper bound on surprisal (Beal, 2003; Da Costa et al., 2020; Sajid, Ball, et al., 2021). This functional can be decomposed in terms of complexity and accuracy, and its minimisation thus means finding an accurate explanation for sensory observations that incurs the least complexity cost:

$$F = \underbrace{D_{KL}[Q(s) \| P(s)]}_{Complexity} - \underbrace{E_{Q(s)}[\log P(o|s)]}_{Accuracy} \quad (1)$$

Here, $D_{KL}$ is the Kullback-Leibler divergence, $o$ and $s$ refer to the outcome and hidden states (or causes), respectively. Free energy depends upon a generative model that comprises a probability distribution $P$ that describes the joint probability of (unobserved) causes and (observed) consequences. This generative model is usually specified in terms of a (likelihood) mapping from hidden causes to outcomes and priors over the hidden causes. The approximate posterior distribution Q in (1) expresses the (posterior) probabilities of hypotheses about hidden states, based on the agent's observations. Uncertainty about anticipated observations is reduced by selecting policies (i.e., probable action trajectories; $\pi$) that *a priori* minimise the expected free energy ($G$)[4] (Parr & Friston, 2019):

$$G(\pi, \tau) = \mathbb{E}_{P(o_\tau|s_\tau)Q(s_\tau|\pi)}[\log Q(s_\tau|\pi) - \log P(o_\tau, s_\tau)] \quad (2)$$

Where $\pi$ refers to a policy, $\tau$ is a (future) time-step. The expected free energy equips the agent with a formal way to assess different policies in terms of how likely they are to fulfil an agent's preferences and information gain about the hidden states of the world. A policy is then selected based on the expected free energy of each policy, which is modulated by the precision parameter $\gamma$:

$$Q(\pi) = \sigma[-\gamma G(\pi)] \quad (3)$$

Thus, the higher the value of $\gamma$, the more precise beliefs about actions. In other words, policy selection becomes more confident. In summary, active inference dictates that (variational and expected) free energy is minimised under a particular model of the environment i.e., a generative model (Friston et al., 2017). These generative models encode particular hypotheses about the current states of affairs. Practically, the model is realised as a partially observable Markov decision process (POMDP) with the assumption that discrete outcomes are caused by discrete hidden states – for technical details see (Da Costa et al., 2020). These models describe the statistical nature of the environment in terms of probability distributions:

---

[3] Surprisal is the negative logarithm of an outcome probability, i.e., $-\ln P(o)$.

[4] For the technical reader, note the resemblance to the terms in Eq.1, but the supplementation of the expectation under the approximate posterior with the likelihood, resulting in the following predictive distribution: $\tilde{Q} = P(o_\tau|s_\tau)Q(s_\tau|\pi)$. This treats planning as inference: Attias, H. (2003). Planning by Probabilistic Inference. Proc. of the 9th Int. Workshop on Artificial Intelligence and Statistics, , Botvinick, M., & Toussaint, M. (2012). Planning as inference. *Trends Cogn Sci.*, *16*(10), 485-488. : i.e., we can evaluate plausible policies before outcomes have been observed.

$$P(o,s,\pi,A,B \mid \zeta,\omega,\gamma) = P(\pi \mid \gamma)P(A \mid \zeta)P(B \mid \omega)$$
$$\times \underbrace{P(s_1)\prod_{\tau=2}^{T} P(s_\tau \mid s_{\tau-1},\pi,B,\omega)}_{Transitions} \underbrace{\prod_{\tau=1}^{T} P(o_\tau \mid s_\tau,A,\zeta)}_{Likelihood} \quad (4)$$

The $A$ parameter encodes the probability distribution of state-outcome pairs (i.e., likelihood distribution), and $B$ encodes the probability distribution of hidden states transitions (i.e., the transition distribution). Both are specified as categorical distributions. Precision terms $\zeta,\omega,\gamma$ are inverse temperature parameters. With high precision, the category with the highest probability converges to 1, whereas for low precision, categories tend to have equal probability (Parr & Friston, 2017; Sajid, Parr, Gajardo-Vidal, et al., 2020). The above probability distributions describe transitions between states in the environment that generate outcomes. Their transitions depend on actions, which are sampled from the posterior beliefs over the policies. Consequently, the sampled actions change the state of the world, giving rise to new outcomes; and continuing the perception-action loop. For the purposes of this paper, we will assume priors over the precision parameters are themselves infinitely precise.

## Precision modulation

We posit that these precision parameters ($\zeta,\omega,\gamma$) can independently modulate bistable perception, since they can shape perceptual confidence and the frequency with which the inferred state of the world alternates. $\zeta$ is the sensory precision over the probabilities of the likelihood distribution $A$ in the generative model, where (hidden) states map onto observations. Thus, sensory precision expresses the confidence with which the model can infer a cause from observations. Practically, high precision (e.g., $>16$) ensures the model can be confident that a particular outcome will be generated reliably by the latent state. Conversely, low precision (e.g., $< 0.2$) implies an ambiguous relationship between causes and outcomes—and observations do little to resolve uncertainty about their causes. The probabilistic mapping from the current state $s_\tau$ to the next $s_{\tau+1}$ is denoted by the state transition matrix $B$. The term $\omega$ encodes the precision of the state transition matrix and it expresses the confidence with which the model can predict the present from the past and *vice versa*. Precision over beliefs about policies is encoded by $\gamma$, which corresponds to the models' ability to confidently select the next action.

We hypothesized that the increase of all three precision terms would lead to a decreased perceptual transition frequency. Furthermore, we hoped to address how to distinguish the influence of each precision term (i.e., neuromodulators) on bistable perception via frequency of eye movements, and acuity (measured using post-switch perceptual confidence). And, finally, via the modulatory effects on neuronal responses encoding distinct percepts of the Necker cube.

## Precision and neuromodulatory systems

These precision parameters have previously been associated with specific neuromodulatory systems (Parr et al., 2018; Parr & Friston, 2017; Sajid, Friston, et al., 2020) – see (Table 1). Briefly, sensory precision ($\zeta$), state transition precision ($\omega$) and policy precision ($\gamma$) can be read as cholinergic, noradrenergic, and dopaminergic neurotransmission, respectively. Some empirical studies suggest a link between the cholinergic release and (the frequency of) perceptual transition. For example, (Sheynin et al., 2020) demonstrate that enhanced potentiation of acetylcholine (ACh) transmission attenuates perceptual suppression during binocular rivalry. Similarly, increased noradrenergic release has also been associated with an altered frequency of perceptual fluctuations (Eienhauser et al., 2008; Pfeffer et al., 2018). (Pfeffer et al., 2018) demonstrate that high catecholamine levels altered the temporal structure of intrinsic variability of population activity and increased the frequency of perceptual alternations induced by ambiguous visual stimulus. Finally, dopaminergic release has also been associated with faster perceptual transition frequency (Schmack et al., 2013).

# Simulations of Necker cube paradigm

In the remaining sections, we model bistable perception, and the intricate precision balance that undergirds it, using simulations of the Necker cube paradigm e.g., (Choi et al., 2020).

## A generative model of the Necker cube

Our generative model of the Necker cube paradigm has two hidden states: *fixation point* and *orientation*, and two outcome modalities: *where* and *feature* (Figure 1). The hidden state *fixation point* has three levels representing bottom-left, top-right, initial position fixation locations, and the *orientation state* has two levels representing left and right orientation. These fixation point locations are motivated by (Choi et al., 2020), who observed eye movements between these particular fixation points during perceptual switches. The outcome *where* reports the location of the eye-fixation: initial, top-right or bottom left. The outcome *feature* reports the corner of the cube being observed: Corner 1 (C1), Corner 2 (C2) or neither (labelled as null).

The likelihood function maps states to outcomes (i.e., state-outcome pairs). Here, the *feature* likelihood is dependent on both *fixation point* and *orientation* factors. For the generative process (i.e., the process we used to generate the observations during simulation), the *where* likelihood depends only on the *fixation point* factor. Therefore, it generates outcomes independently of the *orientation* state. Conversely, the generative model's the *where* likelihood depends on both *fixation point* and *orientation* factors and explicitly maps each fixation point to a specific orientation (see Figure 1). Thus, the bottom-left (top-right) fixation location is only plausible under left (right) orientation. Next, we equipped the model with control states (i.e., states whose transition depend on actions) over eye movements via the *fixation point* factor. Thus, it can control whether to fixate over the top-right, bottom-left, or initial *fixation point*. The *orientation* transition is not controllable and the mapping between current and future states was expressed such that the left (right) orientation always transitions to the left (right) orientation. (Figure 1).

Furthermore, the agent was equipped with strong preferences (measured in nats, i.e., natural logarithm) for avoiding the initial *where* location (-20 nats). This was to encourage the agent to sample bottom-left and top-right locations – as the eye movements between these locations have been shown to be associated with perceptual transitions in the Necker cube paradigm (Choi et al., 2020). At each timepoint, the agent could choose from 3 different actions (i.e., 1-step policy) of either fixating at the initial, bottom-left or top-right location. The prior beliefs about the initial states were initialised to 0.5 for the left and right orientations, 1 for the initial *fixation point* and zero otherwise.

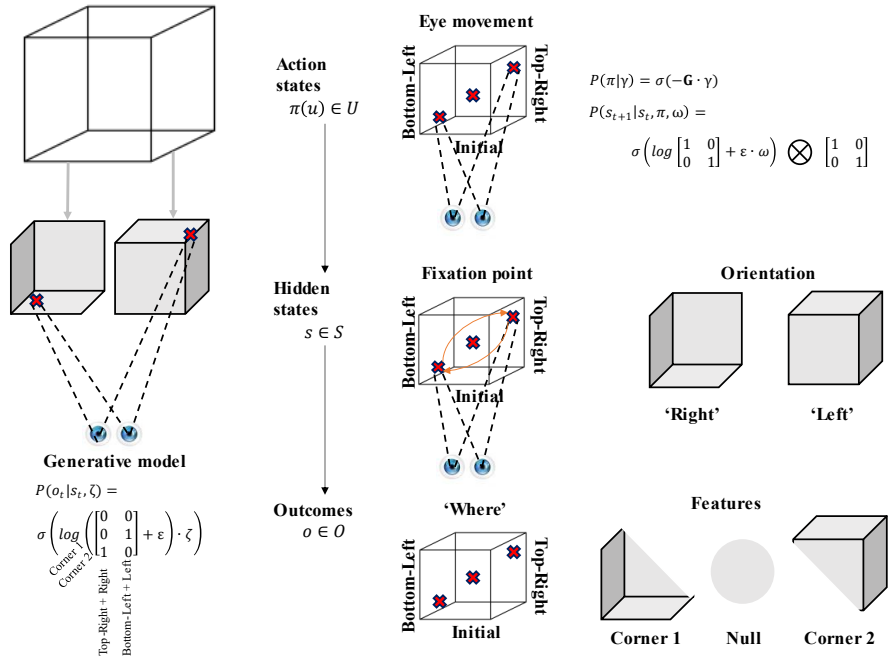

Figure 1. *A graphical representation of the Necker Cube model.* The figure provides a graphical illustration of the generative model with two hidden state factors and two outcome modalities. The first hidden state factor, the *fixation point*, has three levels: bottom-left, top-right, and initial fixation. The second hidden state factor, *orientation*, has two levels, right and left orientation. The outcome modality *features* has three levels: 'Corner 1' (C1), 'Corner 2' (C2), and Null. Here, C1 and C2 denote the two opposite corners and their surrounding areas, and the Null outcome is only plausible under the initial *fixation* point at the first-time step. There is an identity mapping from *fixation point* hidden state factor to *where* outcomes. The likelihood function of the generative model i.e., the probability of an outcome given a hidden state, is encoded such that i) the bottom-left fixation point is more informative about the right orientation as the agent perceives the related C1 corner, ii) the top-right fixation point is more informative about left orientation as the agent perceives the related C2 corner there, and iii) the initial fixation mapped onto a null outcome (i.e., neither C1 nor C2). The *fixation point* transitions (i.e., representing the state transitions across time) are completely precise. This encodes the eye movements between different fixation locations. Conversely, *orientation* transitions for the generative process are non-controllable and transition to the same orientation over time. Here, $\epsilon = e^{-8}$ is a small number that prevents numerical overflow.

## Precision and perceptual alternation

The Necker cube generative model was used to demonstrate the computational mechanisms that underwrite bistable perceptual. For this, we simulated 729 models with different combinations of the three precision parameters: sensory precision ($\zeta$), state transition precision ($\omega$) and policy precision ($\gamma$). The precision values used are specified in Table 2.

$\zeta$ is the sensory precision associated with the likelihood distribution $A$ i.e., which (hidden) states gave rise to particular observations (where $\epsilon = e^{-8}$ is a small number that prevents numerical overflow)

$$A_{i,j,k} = \begin{cases} \sigma(\zeta \cdot log(A_{i,j,k} + \epsilon) \; if \; j \neq k \\ \sigma(0 \cdot log(A_{i,j,k} + \epsilon) \; \text{otherwise} \end{cases} \quad (5)$$

Where *i* represents the outcomes and *j* and *k* represent the *orientation* (either left or right) and *fixation point* (either bottom-left or top-right) factors, respectively. Note, we have excluded the initial *fixation point* for clarity, as its likelihood matrix is uninformative in the generative model. The two factors are unequal either in combination of the bottom-left fixation point and the right orientation or the contrary (see Figure 2). The bold **A** represents the likelihood matrix of how the data are generated (i.e., precise mappings from states to *where* and *feature* outcomes)[5]. Here, the precision parameter $\zeta$ modulates only the columns for the preferred orientation under a given fixation point (i.e., bottom-left fixation point (labelled as 1) maps to the right orientation (labelled as 2) and vice versa), while the unpreferred orientation is parameterised as a uniform distribution. Adjusting the columns of the likelihood matrix in this way can be regarded as manipulating the relative sensitivity of neuronal populations — encoding the probability of each possible (hidden) state to sensory afferents — during model inversion or perceptual inference.

Figure 2A provides a graphical illustration of how the precision parameter values modulate the *feature* likelihood. Here, the sensory precision parameter ($\zeta$) modulates the mapping from orientation states to feature outcomes as a function of location states. Under this parameterisation, a high sensory precision $\zeta \geq 0.5$ (right panels in Figure 2A), leads to a precise likelihood mapping for the state pairs bottom left location – right orientation and top right location – left orientation. Thus, the agent would attribute C1 to the right orientation under the bottom-left position, and C2 to the left orientation under the top-right position. Conversely, under a low sensory precision, the likelihood mapping from an orientation and location to feature outcomes becomes imprecise (left panels in Figure 2A). With this mapping, the agent could not disambiguate between the causes of C1 and C2 outcomes via the perceived orientation regardless of the sampled fixation position. We motivate our choices for these likelihood mappings based on the degree of visibility of the features, assuming that the cube is opaque. Under this assumption, one should not be able to see Corner-1 for a left-oriented cube. Similarly, one should not be able to see Corner-2 for a right-oriented cube (see Figure 1 for left and right orientations). These assumptions are translated as likelihood mappings over the feature outcomes for the aforementioned orientation and fixation point combinations, whose precision is encoded by $\zeta$.

The probabilistic mapping from the current state $s_\tau$ to the next $s_{\tau+1}$ is denoted by the state transition matrix $B$. The term $\omega$ encodes the precision of the state transition matrix in the same fashion as the term $\zeta$:

$$B = \sigma(\omega \cdot log(B + \epsilon)) \quad (6)$$

Where the bold **B** represents the transition of how the hidden states change and give rise to new observations, which is set to be completely precise in the generative process. The B matrix expresses the confidence with which the model can predict the present from the past and the future, and vice versa.

Figure 2B provides a graphic illustration of how precision changes the *orientation* state transition matrix. An increase in the precision of *orientation* state transitions ($\omega$) leads to a precise mapping between the *orientation* at the current and next timepoints (right panel of Figure 2B). With a precise transition matrix, the agent would expect the orientation remain the same over time. Conversely, under a low precision, the agent would expect the orientation to change frequently (left panel of Figure 2B). The modulation of the γ parameter is omitted from this figure, as the best understanding of how this precision parameter influences bistable perception is provided in Equation 3. This parameter modulates the confidence over eye movement selection. If low, this precision prompts the agent switch between the two locations with greater randomness.

---

[5] Here, e(-8) prevents numerical overflow.

Table 2. Precision (hyper-) parameters used to simulate bistable perception. These range from high to low precision values.

| Precision Parameter | Values |
|---|---|
| Sensory $\zeta$ | 0.001, 0.01, 0.1, 0.2, 0.5, 1, 2, 5, 10 |
| State transition $\omega$ | 0.001, 0.01, 0.1, 0.2, 0.5, 1, 2, 5, 10 |
| Policy $\gamma$ | 0.001, 0.01, 0.1, 0.2, 0.5, 1, 2, 5, 10 |

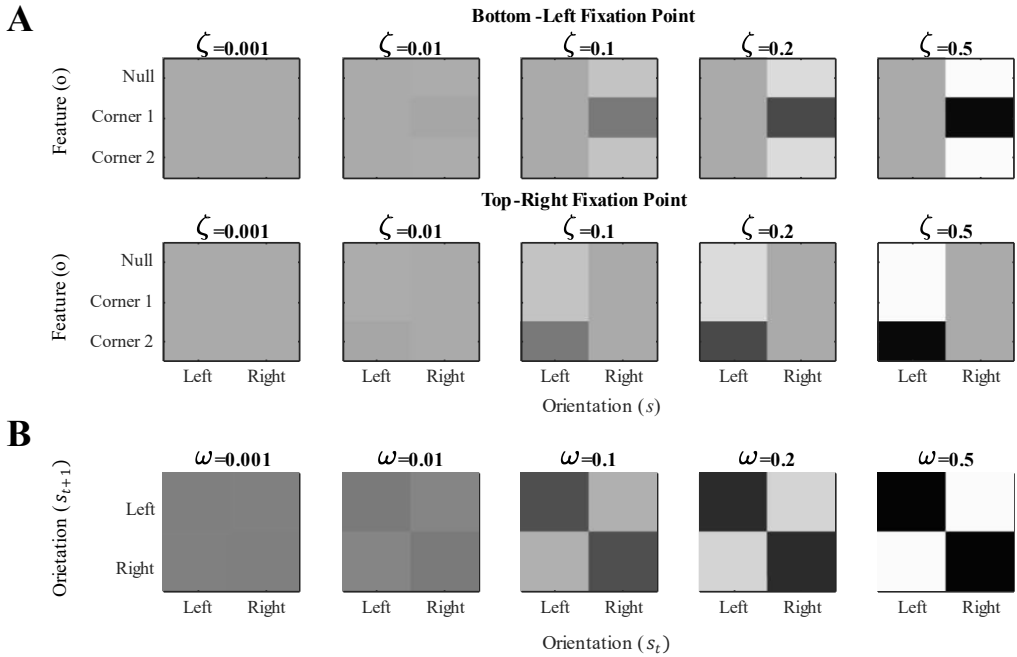

Figure 2. *A graphical illustration of how different precision values change the likelihood and priors of the generative model*. A). A modulation of the likelihood matrix via the sensory precision ($\zeta$). Each row is for a different fixation point with bottom-left on the first and top-right on the second, where the x-axis represents the *orientation* states and the y-axis the *feature* outcomes. B) This panel shows how the state transition precision ($\omega$) perturbations influence the categorical probability distribution of the *orientation* transition. The x-axis represents the *orientation* states at the current timepoint ($t$) and the y-axis the *orientation* states at the next time point ($t+1$). Here, low $\omega$ values lead to a flat distribution which limits the capacity to project current beliefs about *orientation* states to past and future epochs whereas with high $\omega$ the state transition matrix becomes more precise and the capacity to pass messages between epochs increases. For all plots, the scale goes from white (low probability) to black (high probability), and grey indicates gradations in-between. The key difference to note is how the probability distribution shifts from imprecise to precise mappings as we move from low precision values (e.g.,

$\zeta, \omega = 0.001$) to high precision values (e.g., $\zeta, \omega > 0.5$). Values above 0.5 look visually the same as the value of 0.5 and therefore are excluded from this representation.

**Perceptual switch definition**

Next, we quantified what constituted a perceptual switch. This is necessary for quantifying the number of perceptual transitions given particular precision combinations. Here, a switch is counted when a particular orientation (e.g., left) has a high posterior probability ($> 0.5$) at the current time point ($t$) but a low posterior probability ($< 0.5$) at the previous time point ($t-1$):

$$\text{Switch} = \begin{cases} 1 & \text{if } \begin{array}{l} s_{left,t} > 0.5 \text{ \& } s_{left,t-1} < 0.5 \\ s_{right,t} > 0.5 \text{ \& } s_{right,t-1} < 0.5 \end{array} \\ 0 & \text{otherwise} \end{cases} \quad (7)$$

In Equation 7, the bold **s** variables are the probabilities that parameterise our approximate posterior $Q(s)$. Intuitively, this means that a switch is defined as a change from a belief that the left (or right) orientation is most likely to a belief that the right (or left) orientation is most likely.

# Results

## Face-validation

Here, we present a numerical simulation that establishes the face validity of the Necker Cube generative model. For this, we simulated the model with arbitrary precision values; specifically, $\zeta = 0.1, \gamma = 1$ and $\omega = 0.1$ (Figure 3). We observed alternating inferences over the orientation as the trial progressed. This was induced by shifts in eye movements that sampled different corners of the Necker cube. Under our definition, a perceptual switch is observed at time point 7, when both conditions outlined above are met (first row of Figure 3). Conversely, perceptual switch would not be counted at timestep 2 because the posterior probability over the appropriate orientation at the previous timestep 1 is not $< 0.5$ but exactly 0.5. Furthermore, this switch is usually accompanied via an action – see middle panel of Figure 3.

Bistable perception, precision control and neuromodulation

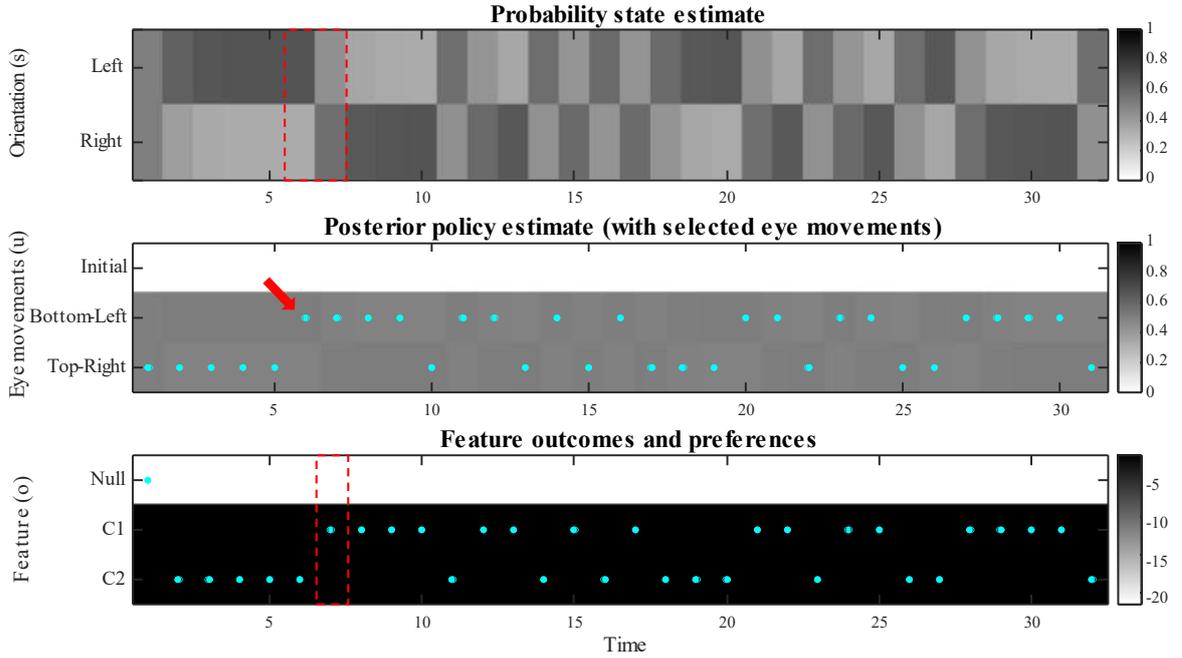

Figure 3. *An example trial with 32 time-steps*. The first row represents the posterior probability for the hidden state orientation. The second row shows which actions, i.e., eye movements, have been selected (cyan dots) and the posterior probability of each policy. This has only 31 time-steps as actions are modelled for the next step. The last row depicts the sampled outcomes over time with cyan dots and the preferences over outcomes with different shades in the background. Here, the light and dark shades illustrate that the agent has a strong aversion for the *Null* outcome observed only at the initial fixation (IF) point but has a relatively higher preference for the C1 and C2 outcomes observed at the bottom left and top right locations, respectively. A perceptual switch is highlighted using the red boxes, where the red arrow in the second row shows that the switch is (mostly) accompanied with an action towards the preferred fixation point. The red box in the last row shows that observing the outcome C1 facilitated the perceptual switch from the left to the right orientation in this instance, as shown in the first row. The example simulation is for the following precision combination: $\zeta = 0.1, \gamma = 1$ and $\omega = 0.1$.

## Simulating perceptual switches

Using the criteria in Equation 7, we measured the number of perceptual switches under different precision combinations (Table 2). Each precision combination was simulated 64 times, using random seed initialisation, with a trial length of 32 epochs. Figure 4 presents the average number of switches under each precision combination. On average, an increase in precision (regardless of the corresponding model parameter) decreases the number of perceptual transitions independently of other precision terms. For example, as $\omega$ increased from 0.001 to 10 we observed a decrease in the number of perceptual transitions. This is unsurprising given our observation regarding Figure 2, i.e., beliefs across time are propagated more confidently for high $\omega$ values. Thus, the orientation does not change frequently during the trial and reduces perceptual switches. For $\zeta$, the increased precision gives higher confidence about what is being perceived, thus removing the uncertainty minimising behaviour that would lead to sampling the other fixation point, which could increase the chances of a perceptual switch.

It is worth noting, however, that for specific combinations of high ζ and low ω values there is an increase in the number of switches. We investigate this in the next section (Figure 4; upper left and middle figures). Lastly, decreased precision over policy selection ($\gamma$) increases perceptual switches. This is because low $\gamma$ values make all policies more likely, leading to a higher frequency of eye movements, and eventual perceptual switch.

The observations above all rest upon a relatively simple insight. For non-zero precision parameters, the best action is always to continue to fixate the same location. This is because the observation associated with our current fixation location supports a belief in a specific orientation (e.g., right orientation if looking at lower left). Under this belief, the alternative location (e.g., upper right) is uninformative as, if the cube were (partially) opaque, there would be little useful visual information there with the opposing corner obscured by the near surface of the cube. In other words, if I am looking to the lower left and infer that the cube is in the right orientation, I would expect that the corner in the upper right will not be visible, so have no reason to look there. As such, the expected free energy will always be lower for the current location compared to alternatives. The result is low switching frequency, with switches that occur only when the action is sampled from the relatively improbable action of moving our eyes. However, the relative improbability of this action is modulated by the precision parameters. Increases in uncertainty about the orientation (via decreases in the sensory[6] or transition precision) attenuate the differences between the expected free energy of each action, resulting in more uncertainty in action selection and increasing the number of switches. Decreases in the policy precision attenuate the influence of the expected free energy on action selection, thus making the improbable action relatively less improbable and increasing switching rate. In short, changes in switching rates occur when greater uncertainty favours more stochastic deviations from an optimal policy of maintaining fixation.

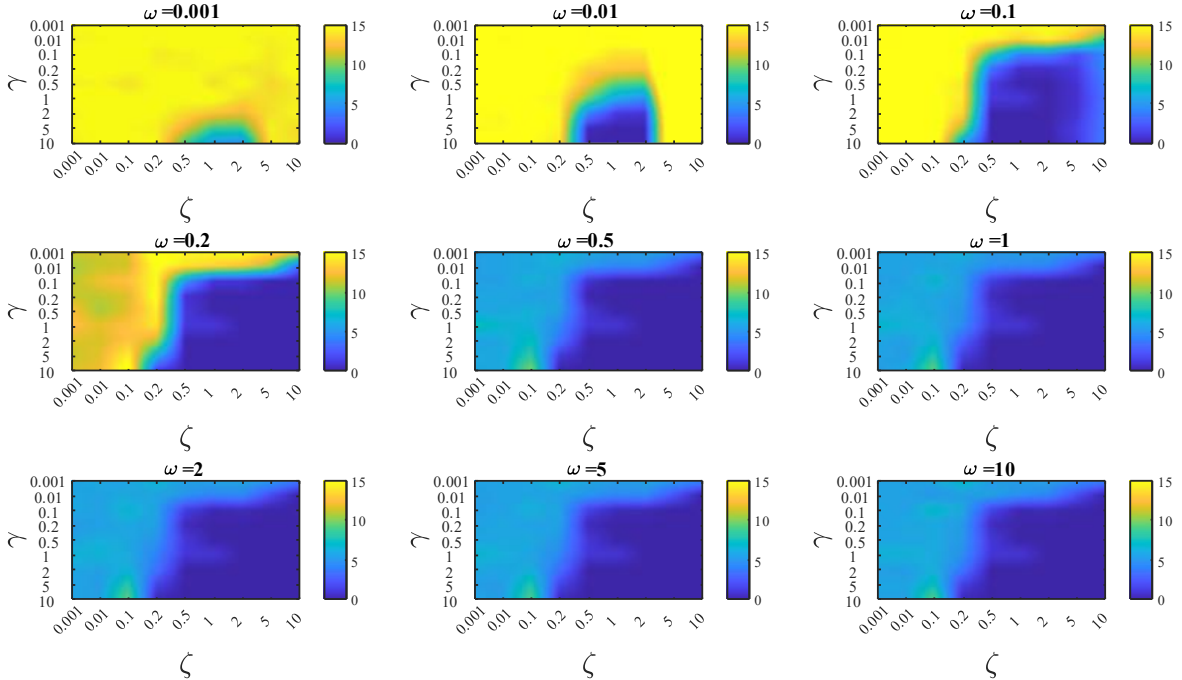

Figure 4. *The average number of switches for different precision combinations.* We plot the average number of switches across 32 trials – each comprising of 32 time-steps. Each heatmap is associated

---

[6] The situation is slightly more complicated for the sensory precision parameter, as this has a dual role. The first is in determining the confidence in the orientation (as inferred from potentially unreliable sensory data). The second is in determining the ambiguity (which contributes to the expected free energy) of each location.

with distinct $\omega$ values. The x-axis is associated with $\zeta$ and y-axis plots the different $\gamma$ value. The average switch count ranges from 0 (dark blue) to 15 (yellow).

## Dissociating individual precision manipulations

To dissociate the individual influences of each precision during bistable perception, we investigated the (average) posterior probability of the cube's orientation after the switch occurred – alongside average switch rates. Here, the posterior probability denotes the $s_{left,t}$ or $s_{right,t}$ value used to identify a switch (Equation 7). The differences across each precision were evaluated by considering each individually and taking its (marginal) average across all possible combinations (Figure 5; Table 3). These differences revealed that posterior switch probability and average switch rate for both $\zeta$ and $\omega$ followed a non-linear relationship – as modelled with a polynomial expansion (Table 3). Conversely, we observed a negative linear association (Table 1; i.e., 1st-order polynomial) for the posterior switch probability and average switch number for γ. This highlights that high posterior switch probability (i.e., values $> 0.5$), that determines switch rate, can manifest in multiple ways – see supplementary text for further analysis. Furthermore, there is a degenerate (many to one) mapping between the switch posterior probability and number of switches across the different precision terms (Figure 5). This speaks to the multiple, but distinct routes through which perceptual transitions can arise.

Table 3. Fitted polynomial coefficients across different precision values for posterior switch probability (A) and average switch number (B). The relationship between precision and perceptual switching was modelled with the polynomial expansion: $y = \beta_0 + \beta_1 x + \beta_2 x^2 + B_3 x^3 + \varepsilon$, and its fit was measured using sum of squares of errors (SSE). Here, * denotes 10% significance level, ** denotes 1% significance level and *** 0.1% significance level.

|  | $\beta_0$ | $\beta_1$ | $\beta_2$ | $\beta_3$ | SSE |
|---|---|---|---|---|---|
| **A. Posterior switch probability** ($y$) | | | | | |
| $\zeta$ | 0.3837*** | 0.1005** | $-0.0069$* | - | 0.0030 |
| $\omega$ | 0.7840 | $-0.1308$ | 0.0277 | $-0.0017$ | 0.0053 |
| $\gamma$ | 0.6658*** | $-0.0069$*** | - | - | 0.0003 |
| **B. Average switch rate** ($y$) | | | | | |
| $\zeta$ | 7.5820 | 2.9380 | $-1.0210$ | 0.0740 | 7.1036 |
| $\omega$ | 19.6500*** | $-4.4870$** | 0.2955* | - | 6.9587 |
| $\gamma$ | 9.1480*** | $-0.5155$** | - | - | 3.3096 |

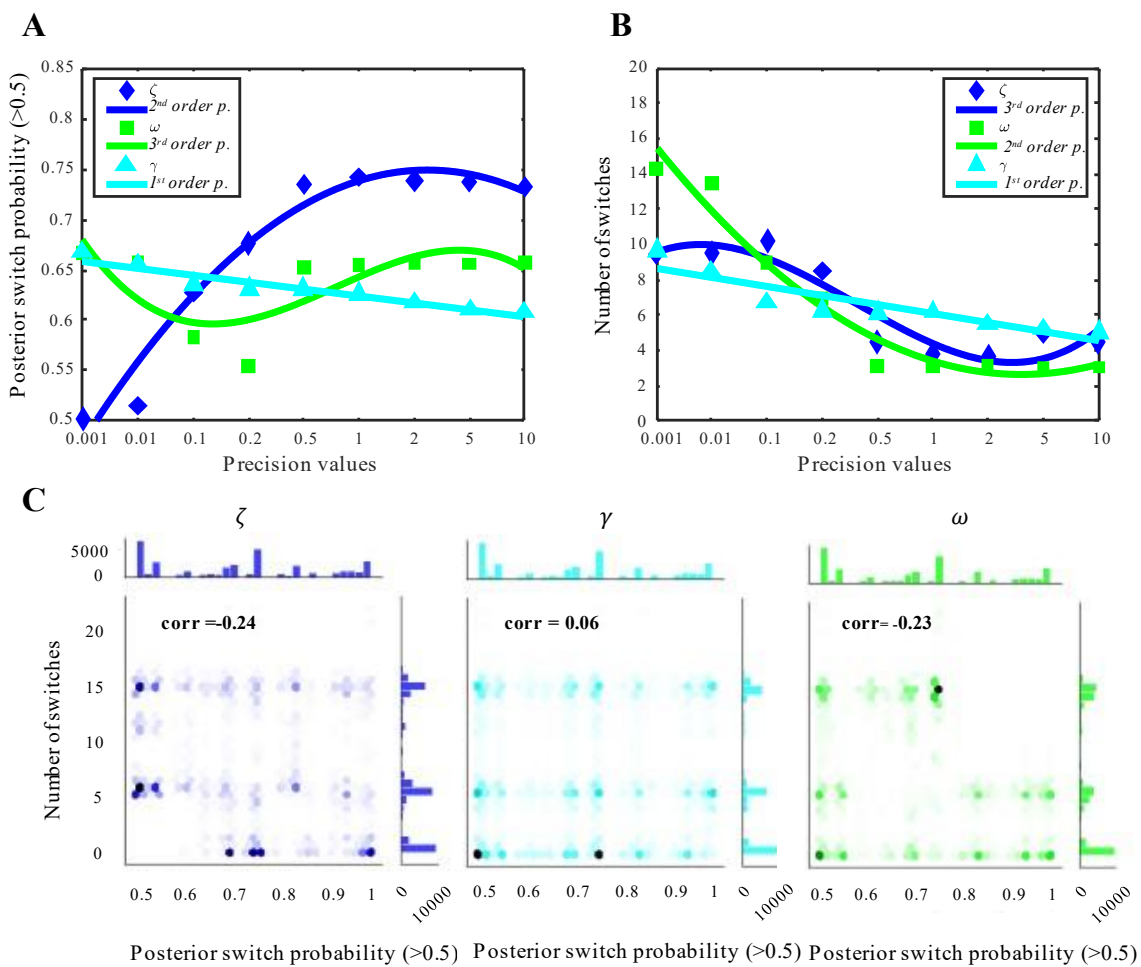

Figure 5 *Dissociating individual precision terms*. For **A** and **B**, each data point represents the average switch posterior probability (**A**; y-axis) and the number of switches (**B**; y-axis) across different precision values (x-axis). The curves represent the fitted polynomials for each precision value: ζ (blue), ω (green) and γ (cyan). **C**) the joint-plot of the association between number of switches and posterior switch probability. The x-axis presents the posterior switch probability, y-axis the number of switches. Here, each plot presents a different precision term.

## Discussion

We investigated how precision manipulation could underwrite bistable perception. For this, we cast bistable perception, the phenomenon where perception alternates between distinct interpretations of a static stimulus, as an enactive process associated with specific eye movements that shift the focus from one visual feature to another leading to a perceptual transition (Choi et al., 2020). This ensues from a dissociation between the inferred percept and sensory observation (Brascamp et al., 2018) as distinct features of the visual stimulus are sampled. Computationally, we show that the frequency of switches between the two percepts depends on a modulation of (at least) three precision terms that determine the confidence of posterior beliefs. Here, we illustrated that there are distinct ways in which precision (hyper-) parameters – associated with neuromodulators – can interact to affect bistable perception and how their influences can be dissociated from each other using post-hoc analysis of posterior beliefs.

Below, we relate distinct precision terms to neuromodulators based on the previous literature review (Parr & Friston, 2017).

**Precision manipulation and neuromodulation**

Sensory precision is thought to be modulated via acetylcholine in the active inference framework (Parr & Friston, 2017) and in normalization models (Schmitz & Duncan, 2018). The influence of this neuromodulator on bistable perception has been studied in (Pfeffer et al., 2018; Sheynin et al., 2020) with apparently inconsistent results of either no influence on the switching rate or decreasing it, respectively. Based on our analysis, we found that sensory precision $\zeta$ depends on other precisions when it comes to the switching frequency (Figure 4) and so looking at the switching rate alone seems insufficient to dissect the specific contribution of this neuromodulator. For this reason, we fitted our simulated data to polynomial expansions to disentangle contribution of individual precision terms. From this analysis, we see that the increase of sensory precision should accentuate the acuity of perceived orientation—assumed to be equivalent to the post-switch perceptual confidence—which is consistent with (Sheynin et al., 2020).

The $\omega$ precision has previously been associated with noradrenergic release (Parr & Friston, 2017). A study by (Pfeffer et al., 2018) used a noradrenaline reuptake inhibitor to study this. They found that after administering a drug boosting noradrenaline, participants reported a faster switching rate of a bistable stimulus. As stated above, it is difficult to dissect a specific contribution of neuromodulators (considering them as precision modulators) in bistable perception tasks given only the measure of switching rate. Moreover, bistable perception shows a close link to pupil dilation (Einhauser et al., 2008; Hupé et al., 2009; Kloosterman et al., 2015) which is linked to noradrenergic release (Larsen & Waters, 2018), and so future work could target pupil dilation in addition to the eye movements in our current model.

We also showed that high policy precision $\gamma$ decreases the frequency with which bistable perception alternates. This precision parameter is suggested to be related to dopamine (Parr & Friston, 2017), but few studies have looked at the role of dopamine and bistable perception or binocular rivalry. Nevertheless, a study by (Schmack et al., 2013) showed that there is an observable alternation of perceptual switches associated with DRD4 gene carriers but this effect was found only for a specific allele (DRD4-2R) but not for others (DRD4-4R and DRD7R). Moreover, (Kondo et al., 2012) found no effect of dopaminergic genes on the rate of visual perceptual switches. However, for auditory bistable perception, presence of prominent alleles for synthesizing this neurotransmitter decreased the number of switches. It is also not fully understood how these specific genes affect the dopaminergic neurocircuitry, thus a specific conclusion on whether and how dopamine targets bistable perception is still open.

**Neuroanatomy**

The deployment of the precision terms studied here can be associated with feature-based attention (FBA), as the perceptual switches here are understood as switches of orientation. This view is also corroborated with a similarity of brain regions involved in processing bistable perception and FBA, as both activate regions such as frontal eye field, intraparietal cortex, temporoparietal junction, and inferior frontal junction (Brascamp et al., 2018; Loued-Khenissi & Preuschoff, 2020; Zhang et al., 2018). Interestingly, all the neuromodulators suggested to be related to distinct precision terms used here are involved in attentional processing (Thiele & Bellgrove, 2018). It is possible that the FBA network deploys attentional mechanisms partially by regulating distinct neuromodulators that lead to distinct neurobiological changes but to overlapping behaviours. This relates to the previously reported top-down modulation of bistable perception via the fronto-parietal network (De Graaf et al., 2011).

**Limitations and future directions**

A key limitation of our work is that we included a limited amount of possible fixation points which makes the study of eye movements over-simplified. Including more fixation points – and thus actions – could provide a more applicable model for empirical studies and fitting of real data. Next, we pre-specified the initial probabilities and precision values instead of updating them during each trial. Future work should explore online selection of these precisions and how they influence bistable perception based on a task. This is related to mental actions i.e., internal process that change posterior beliefs by regulating precision (Limanowski & Friston, 2018; Metzinger, 2017) – and can be added via a hierarchical model in which slower parts of the model modulate the precision terms that influence faster dynamics (Hesp et al., 2021). Understanding how the precision parameters are learned, we could also examine the dynamics of neuromodulators, as so far, we have studied these effects in a stationary environment.

# Conclusion

We have shown how bistable switches can be manipulated via three distinct precision terms. Moreover, we disentangled among their influences, using changes in posterior beliefs to identify perceptual switches. The remaining question concerns the plausible implementation of these precision terms in the brain, which is currently suggested to be related to cholinergic, noradrenergic and dopaminergic neurocircuitries to state transition, likelihood, and policy selection precision terms, respectively. Overall, our results speak to a degenerate functional architecture that supports the switching rate of bistable perception (Noppeney et al., 2004; Price & Friston, 2002; Sajid, Parr, Hope, et al., 2020) i.e., multiple neuromodulatory systems can modulate the perceptual switching rate.

**Software note:**

The generative model in these kind of simulations changes from application to application; however, the belief updates are generic and can be implemented using standard routines (here spm_MDP_VB_X.m). These routines are available as Matlab code in the SPM academic software: http://www.fil.ion.ucl.ac.uk/spm/. The code for the simulations presented in this paper can be accessed via https://github.com/filipnovicky/BistablePerception.

**Funding:** This work was funded by the Serotonin & Beyond programme (953327, FN), the Medical Research Council (MR/S502522/1, NS), Wellcome Trust (Ref: 203147/Z/16/Z and 205103/Z/16/Z, KJF) and a 2021-2022 Microsoft PhD fellowship (NS). KJF was also supported by a Canada-UK Artificial Intelligence Initiative (Ref: ES/T01279X/1).

**Authors Contributions:** All authors made substantial contributions to the conception, design and writing of the article; and approved publication of the final version.

## Supplementary Materials

We looked at the differences between the precision values, for each precision term across each posterior switch probability (Figure S1A) and the number of switches (Figure S1B) – using two-side t-tests. We observed a difference across precision values for $\zeta$ and $\gamma$, but high $\omega$ values ($\geq 0.5$) are indistinguishable. For number of switches, only low and high precision value clusters are distinguishable.

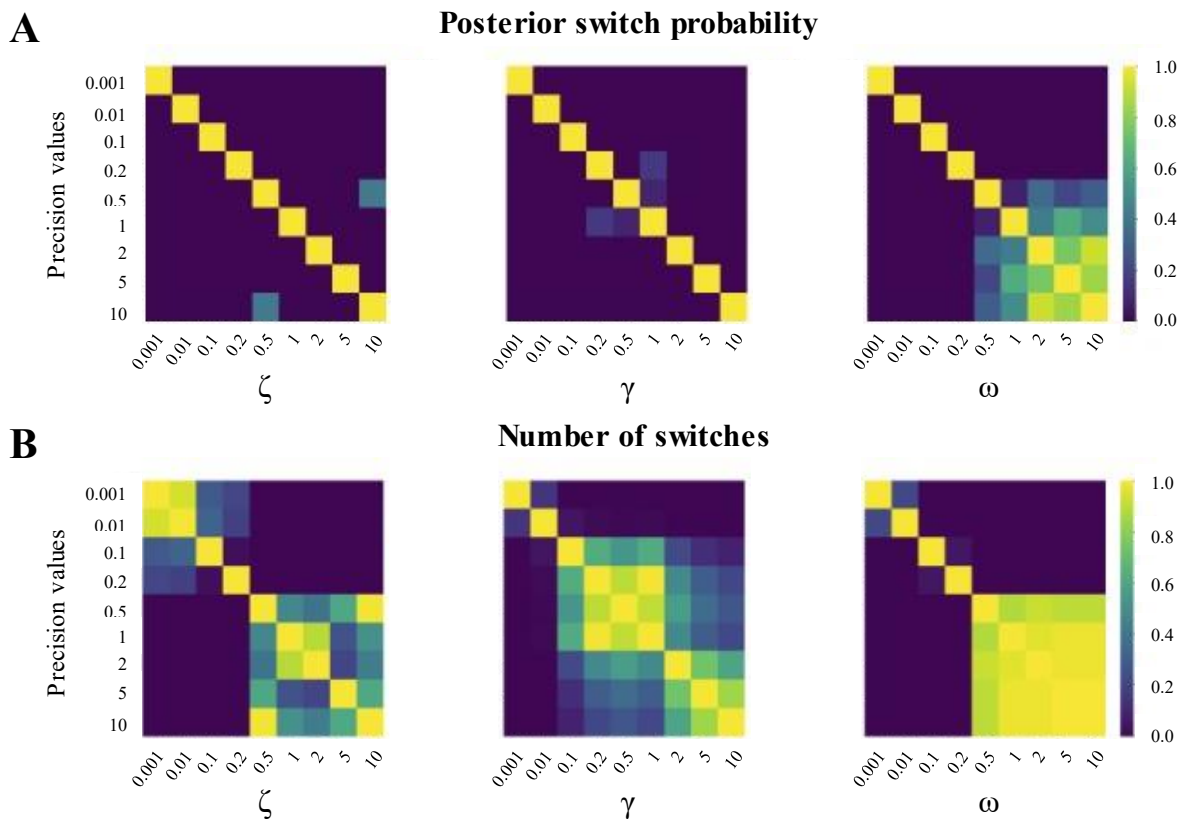

Figure S1: Two-sided t-tests using unequal variance. Top row (A) plots the results for posterior switch probability and bottom row (B) for number of switches. Here, the heatmap plots the p-values for each t-test; lower values are indicative of statistical differences (p-value < 0.0001) and high values (p-values >0.5) represent failure to reject the null hypothesis i.e., the two precision values are the same.

Bistable perception, precision control and neuromodulation